\documentclass[10pt]{article}
\usepackage{amsmath}
\usepackage{graphicx}
\usepackage{amsfonts}
\usepackage{amssymb}
\usepackage{epsf}
\usepackage{latexsym}

\def\be{\begin{equation}}
\def\ee{\end{equation}}
\def\bea{\begin{eqnarray}}
\def\eea{\end{eqnarray}}

\def\brho{\mbox{\boldmath$\rho$}}
\def\bpi{\mbox{\boldmath$\pi$}}

\def\CPI{Conditional Probabilities Interpretation }

\def\beq{\begin{equation}}
\def\eeq{\end{equation}}
\def\bea{\begin{eqnarray}}
\def\eea{\end{eqnarray}}
\def\pa{\partial}
\def\d{\textrm{d}}

\def\5Star{\mbox{\Large$\star$}}

\def\cr{\mbox{\scriptsize{\bf $\mbox{ } \times \mbox{ }$}}}

\def\sumi3{\sum\mbox{}_{\mbox{}_{\mbox{\scriptsize $i$=1}}}^3}

\def\sumj3{\sum\mbox{}_{\mbox{}_{\mbox{\scriptsize $j$=1}}}^3}
\def\sumk3{\sum\mbox{}_{\mbox{}_{\mbox{\scriptsize $k$=1}}}^3}

\def\mB{\mbox{B}}

\def\mF{\mbox{F}}

\def\sA{\mbox{\scriptsize A}} 
\def\sB{\mbox{\scriptsize B}}

\def\sE{\mbox{\scriptsize E}}
\def\sF{\mbox{\scriptsize F}}

\def\sJ{\mbox{\scriptsize J}}

\def\sN{\mbox{\scriptsize N}} 
\def\sO{\mbox{\scriptsize O}}

\def\sS{\mbox{\scriptsize S}}

\def\eph{\mbox{\scriptsize eph}}
 
\def\eph(B){\mbox{\scriptsize em(JBB)}}

\def\eph(B){\mbox{\scriptsize emergent(JBB)}}

\def\sbfM{\mbox{\bf \scriptsize\sffamily M}}

\def\bh{\mbox{{\bf h}}}

\def\bq{\mbox{{\bf q}}}

\def\fM{\mbox{\sffamily M}}
\def\fN{\mbox{\sffamily N}}

\def\fQ{\mbox{\sffamily Q}}

\def\sfA{\mbox{\sffamily{\scriptsize A}}}

\def\sfZ{\mbox{\sffamily{\scriptsize Z}}}
\def\K{Kucha\v{r} }

\def\bB{\mbox{\bf B}}

\def\bQ{\mbox{\bf Q}}

\begin{document}
\vspace{.7in}
\begin{center}
\large{\bf MACHIAN TIME IS TO BE ABSTRACTED FROM $\mbox{\boldmath{$WHAT$}}$ CHANGE?}\normalsize

\vspace{.15in}

{\large \bf Edward Anderson}\footnote{{\em APC, Universit\'{e} Paris Diderot} and {\em DAMTP, Cambridge}, 
edward.anderson@apc.univ-paris7.fr}

\begin{abstract}

``{\it It is utterly beyond our power to measure the changes of things by time. 
Quite the contrary, time is an abstraction at which we arrive through the changes of things.}'' Ernst Mach \cite{M}.  

\mbox{ }  

\noindent  {\sl What} change?  
Three answers to this are `any change' (Rovelli), `all change' (Barbour) and my argument here for the middle ground 
of a `sufficient totality of locally relevant change' (STLRC) giving a generalization of the astronomers' ephemeris time.
I then use STLRC as a selection principle on existing and new approaches to the Problem of Time in Quantum Gravity.
Emergent Jacobi-Barbour-Bertotti time can be interpreted as arising from a STLRC, resolves the classical 
Problem of Time and has an emergent semiclassical counterpart as regards facing the QM Problem of Time.

\end{abstract}

\end{center}

\section{Introduction: Machian time and three different attitudes to it}\label{Sec-1}

{\sl Any} change and {\sl all} change were considered in, respectively, e.g. Rovelli's \cite{Rfqxi} and Barbour's \cite{Bfqxi} 
Foundational Questions Institute contest winning essays of 2008.  
A third answer I offer here is {\it a sufficient totality of locally relevant change (STLRC)}, which gives rise to a {\it generalized local ephemeris time (GLET)}. 
See also e.g. \cite{BB82, B94I, EOT, Rovellibook, ARel} for these three points of view; in the present article I argue in favour of this third perspective.

\mbox{ }  

\noindent I first analyze these in terms of three separate alternatives concerning what constitutes a good clock.

\mbox{ } 

\noindent 1A) Any subsystem suffices as a clock or 1B) some subsystems are better clocks than others.

\mbox{ } 

\noindent 2A) Clocks are localized subsystems or 2B) clocks are to factor in changes of the actual subsystem under study \cite{Clemence}.  

\mbox{ } 

\noindent 3A) The less coupled one's clock and subsystem under study are, the better the clock is, or 3B) the two fundamentally {\sl have} to be coupled.  
This last alternative can be resolved with a compromise: take a small but strictly nonzero coupling, so the subsystem under study only very gently disturbs the clock's physics 
(this additionally fits in with how nothing can shield gravity).  

\mbox{ }

\noindent Perhaps then, time is to be abstracted from {\sl any} change [1A)]. 
I will refer to this as {\bf AMR timekeeping} after Aristotle, Mach and Rovelli, 
since it is similar to `{\it tot tempora quot motus}' - so many times as there are motions - the Scholastics' interpretation of Aristotle (see p 54 of \cite{Jammerteneity}).  
This carries connotations of 2A), which is also widespread in modern Theoretical Physics \cite{PW83GPP}, probably for the reason given in II) below.  

\mbox{ }

\noindent Perhaps instead time is to be abstracted from {\sl all} change in the universe.  
I will refer to this as {\bf LMB timekeeping} since it is Barbour's position \cite{B94I, EOT} and leans on Leibniz as regards the universe itself being the only perfect 
clock.\footnote{`Perfect' here is meant at the classical level. 
See e.g. \cite{52} for various obstructions to the notion of perfect clock at the quantum level (finite probability of quantum clocks running backward, Salecker--Wigner inequalities).}
%
This rests on 1B) and 2B) and the following further principle.

\mbox{ } 

\noindent 4) `The more change the better'. 

\mbox{ } 

\noindent Moreover, in LMB timekeeping, 4) is taken to its logical extreme:
 
\mbox{ }   
 
\noindent 5) `include all the change'.  

\mbox{ }

\noindent Much can be learnt by contrasting the `any' and `all' positions.
5) provides, within the context of 2B), an {\sl incontestable} time, in that there are no 
more changes elsewhere that can cause one to doubt such a time.  
1A) has an underlying sense of `democracy', argued to be useful in generic situations that are crucial in General Relativity (GR), 
by which there would appear to be no privileged time standard. 
But 1B)'s `meritocracy' is in close accord with mankind's history of accurate timekeeping;
both Barbour \cite{Bfqxi} and I invoke this as useful evidence, although it leads each of us to different conclusions.  

\vspace{10in}

\noindent {\bf Sidereal time} is kept by the rotation of the Earth relative to the background of stars, and served 
satisfactorily as a timestandard from Ptolemy \cite{Ptolemy} until the 1890s.  
Barbour  \cite{Bfqxi} pointed out that this reflects 1B), to the extent that by those 2 millennia's standards, it is reliable for 
making predictions about the other celestial bodies, unlike solar time. 
[Moreover, his argument is for {\sl apparent}, rather than mean, solar time, which is far closer to 
sidereal time and was indeed likewise used reliably.]  
Sidereal time served well since the Earth's rotation is stable to 1 part in 10$^{8}$, 
alongside the background of stars only changing very slowly.
It also reasonably embodies 2A) since the Earth is localized in comparison to the solar system.

The main point of discussing sidereal time is, however, its demise due to irregularities in the Earth's rotation.  
This was noted via departures from predicted positions, especially noticeable for the Moon.
As De Sitter said \cite{DS27}, {\it ``the `astronomical time', given by the Earth's rotation, and used in all practical 
astronomical computations, differs from the `uniform' or `Newtonian' time, which is defined as the independent variable of 
the equations of celestial mechanics."} 
This difference was addressed by using the Earth-Moon-Sun system as a superior timestandard provider [1B)],   
to be solved iteratively by Clemence's proposal \cite{Clemence, Alma}.  
This is known as {\bf ephemeris time}.  
It is a prime example of the highly Machian perspectives 2B) and 4) [for all that Clemence did not present it in such philosophical light], 
whilst also avoiding the extreme position embodied by 5).  

\mbox{ }

\noindent Of importance for further understanding the different timestandards in use today \cite{Clemence48}, there is a 
divergence between civil and dynamical timekeeping, due to the Earth's far more central role in the former.
Everyday life is run according to the Earth's physics, despite the small tidal braking on its rotation. 
There have been two subsequent significant changes to timekeeping.      

\mbox{ }

\noindent I) {\bf Relativistic effects} are significant at accuracy of 1 part in $10^{12}$, which became relevant in the late 1970s.  
Thus one ceased to be able to time-keep within Newton's conception of time; 
its relativistic replacement has the novel feature of

\noindent depending on the position and velocity of the clock 
(by GR redshift and SR time dilation respectively).  

\mbox{ } 

\noindent II) {\bf Advent of atomic clocks}. It was these that permitted accuracy to exceed the above figure  
in the late 1970s; by today we have individual clocks with accuracy of 2 parts in $10^{16}$ \cite{LGS11}.
These accuracies exceed those of astronomical timestandards, which are limited by not knowing the detailed composition of solar system bodies. 
This goes a long way toward explaining why astronomical timestandards have become rather less well-known in Theoretical Physics.
In contrast, atomic clocks, have simple and well-known internal constitution and physics, whilst being designed for great stability. 
They are small [2A)], all the better for shielding them from disturbances and being convenient `reading hands' from which to quickly take note of times.

\mbox{ }

However, are atomic clocks a fully independent paradigm, or are they more like far more wieldy reading hands than the position of the Moon, but that, nevertheless, need recalibrating? 
A fundamental issue is that \cite{Schegel, Clemence}, independently of how stable the clock is, why should 
what it it reads out correspond {\sl exactly} to the dynamical time (an SR-and-GR-upgraded counterpart of the Newtonian time) of the system under study? 
As it happens, in the early days of atomic clocks it was established that they read out ephemeris time to at least 1 part in 10$^9$ \cite{Parry}, 
which greatly eased the transition from ephemeris to atomic timestandards. 
However, this has the status of a {\sl null experiment}, so one should keep on testing whether it continues 
to hold true as accuracy improves.

After all, it was along these lines that one took the Earth {\sl not} to read off the dynamical Newtonian time in the upgrade from sidereal to ephemeris time.
Moreover, it helps here to decompose 2B) into two logically-possible problems: with the physics of the clock itself, and with 
the clock-reading not corresponding to the parameter that most simplifies the equations of motion, which are more holistic in depending 
on the system under study rather than the clock.
The Earth is a `dirty clock' due to its complex and partly unknown internal dynamics, 
whereas atomic clocks are designed to be far cleaner. 
This is unless the holistic effect occurs at some level; have we seen any evidence for this yet?  
For sure, the {\it leap seconds} that one adjusts some years by are not of this nature, as they concern 
adjusting {\sl civil} time to compensate for irregularities in the Earth's rotation.  
Further adjustments (3 to 5 orders of magnitude smaller) between `barycentric dynamical time' (for space science) and 
`terrestrial time' (for science on Earth) are fully accounted for by standard relativistic physics.  
To that accuracy, at least, modern astronomical timestandards can be readily obtained by shifting atomic ones, 
so no holistic realizations of 2B) are in evidence.
This is worth pointing out since i) elsewise the revelation that Clemence's work on ephemeris time is conceptually Machian 
might well be taken to carry holistic connotations. 
ii) Nevertheless, one might keep an open mind as to whether holistic connotations will show up at some level much finer 
than the sidereal to ephemeris time transition of the first half of the 20th century, by developing a Machian time program  
{\sl as well as} treating the minutiae of further improving atomic clocks.  

\mbox{ } 

\noindent Taking the above insights on board, a small amount of generalization leads to the following third picture 
for a practical {\it LMB-CA timekeeping} (the C stands for `Clemence'). 
This represents `middle ground' between Barbour and Rovelli's positions whilst being maximally in accord with mankind's history of accurate timekeeping.

\section{Generalized Local Ephemeris Time (GLET) procedure}\label{Sec-2}

The third option considered in this paper is that perhaps {\sl time is best in practice abstracted from a STLRC}.    
\noindent This gives a Generalized Local Ephemeris Time (GLET), which is a timestandard of the LMB-CA type.
I.e., do not just use a change to abstract a time, but also check whether using it in the equations of motion for other 
changes suffices to predict these to one's desired accuracy, in accord with 1B), 2B) and 4).   

\noindent Note that, in comparison to AMR, it has the explicit extra element of acknowledging that some times are 
locally more useful than others.  
For, even in generic situations, one can locally consider a ranking procedure for one's candidate times, alongside a refining procedure 
until the sought-for (or physically maximal) accuracy is attained.  
Whilst these may well give a less accurate local timestandard than that for the highly non-generic Earth-Moon-Sun system, 
say, these should nevertheless still give rise to an extremum.
This establishes that Rovelli's good idea of considering genericity is not solely the province of `any time' approaches.
\noindent Also, in comparison to LMB timekeeping, the ephemeris-type method as used in practice for the solar system is entirely adequate 
without having to consider the net physical effect (tidal effect) of very distant massive bodies as per Sec \ref{Sec-4}.  
\noindent Thus the LMB-CA account of practical timekeeping holds without need for LMB's whole-universe extreme 
5), which is a sensible avoidance for the following reasons. 
Firstly, the details of very distant bodies are but inaccurately known to us.  
Secondly, it would open Pandora's box as regards how to reconcile Leibniz's meaning of `universe' with that of modern GR cosmology.

The GLET procedure thus entails not just using a change to abstract a time, but also checking whether using this time in the equations of 
motion for other changes suffices to predict these to one's desired accuracy.  
If the answer is yes, then we are done.  
If not, consider further locally-significant changes as well/instead in one's operational definition of time 
(locally significant as judged by the criterion in Sec \ref{Sec-4}).  
Then if this scheme converges without having to include the entirety of the universe's contents, one has found a LMB-CA time that is 
locally more robust than just using {\sl any} change in order to abstract a time, and it is particularly useful for  
equations of motion in, and propositions conditional on, this time.  

\mbox{ } 

\noindent {\sl Where} in space the timestandard is supposed to hold matters once accuracy exceeds GR and SR corrections.  
Abstracting time from the Earth-Moon-Sun system does {\sl not} mean finding time throughout some box whose side
is some characteristic length for this system. 
Different objects moving in different ways therein will need specific SR and GR corrections to the timestandard 
if that is to attain a certain level of accuracy.

\section{Jacobi-Barbour-Bertotti approach to the Problem of Time}\label{Sec-3}

The relational reformulation of mechanics \cite{BB82} and GR \cite{RWR, Phan} produces a candidate time function that 
I refer to as the Jacobi-Barbour-Bertotti (JBB) time, $t^{\sJ\sB\sB}$ \cite{B94I, SemiclI, FileR, ARel}. 
This can be interpreted as of LMB or LMB-CA form (see Sec \ref{Sec-4} for this distinction), 
and it clarifies and resolves many parts of the notorious Problem of Time (PoT) in Quantum Gravity.  
The present article argues that the LMB-CA implementation of Mach's `time is abstracted from change' is of particular merit and thus  
should be considered as a possible guide among the many facets of, and strategies for, the PoT.  
The reader should be warned that the PoT is often mistaken for the quantum-level Frozen Formalism Problem of eq (\ref{FFP}).  
In fact, this is but 1 of around 8 facets that the PoT possesses \cite{Kuchar92I93, APoT2}, 7 of which already have classical counterparts in 
GR, and 5 of which are even present in a suitably relational reformulation of classical mechanics for the universe as a whole.                                                                                                                                                                                                                                                                                                                                                                                                                                                                                                                                                                                                                                                                                                                                                                                                                                                                                                                                                                                                                                                                                                                                                                                                                                                                                                                                                                                                                                                                                                       
This classical part of the PoT can be viewed as a tension in the idea of `time for the universe as a whole' \cite{ARel, APoT2}.  
It consists of the following parts.  

\mbox{ }

\noindent 1) {\bf Temporal Relationalism}.  
Following Leibniz \cite{L}, one postulates there to be no time for the universe as a whole.  
{\sl In this sense, there is a {\sl classical} Frozen Formalism Problem right from the outset, from insisting on 
modelling closed universes from a background-independent perspective.}
Leibniz's idea can be mathematically implemented by considering parametrization-irrelevant actions that include no 
extraneous time-like variables such as absolute Newtonian time or the GR lapse \cite{BSW}.  
Such actions are then of Jacobi type \cite{Lanczos} for mechanics or of Baierlein-Sharp-Wheeler type 
\cite{BSW} for GR-as-geometrodynamics (i.e. formulated as a dynamics of 3-geometries).  
For now, I only present the Jacobi reformulation of mechanics, which follows from the action principle 
\beq
S =  \sqrt{2}\int\sqrt{E - V}\,\d s \mbox{ } , \mbox{ } \mbox{ } \mbox{ for {\it kinetic arc element} } \mbox{ } \d s := \sqrt{m_I \d \bq^{I\, 2}}
\eeq
for particle positions $q^{I\mu}$ (with particle index $I$ and spatial index $\mu$, particle masses $m_I$, potential energy 
$V$ and total energy $E$.
Actions of this sort were in use for many years prior to realizing that these are (or are easily recastable as) 
manifestly temporally relational, via the work of Barbour and collaborators \cite{BB82, B94I, RWR, San, FEPI, B11GrybThPooley, FileR, ARel}).

As envisaged by Dirac \cite{Dirac}, parametrization-irrelevant actions oblige the appearance of constraints that are (i.e. solely due to 
the structure of the Lagrangian rather than requiring any kind of variation).  
In the present case, the structure of the Lagrangian gives a single primary constraint that has quadratic and no linear dependence in the momenta.  
For mechanics, this is the energy constraint $H := T + V = E$; in the GR-as-geometrodynamics analogue below, we shall see the Hamiltonian 
constraint arise likewise, whose form is well-known to lead to the {\sl quantum} Frozen Formalism Problem facet of the PoT 
(although it is far less well known to arise by the complete chain of reasoning I emphasize above).

Moreover, one can now resolve the classical Frozen Formalism Problem via a JBB emergent time \cite{B94I, SemiclI, FileR} that {\sl simplifies the momenta and equations of motion}; its form is 
\beq
t^{\sJ\sB\sB} =  t^{\sJ\sB\sB}(0) + \left.\int\d s \right/\sqrt{2\{E - V\}} \mbox{ } .  
\eeq
It attains this simplification by being a relational recovery of the Newtonian time of ordinary Mechanics \cite{B94I}. 
It is an example of Mach's `time is abstracted from change'; on the face of it, it is an LMB implementation since all changes are present in the $\d s$, 
though Sec \ref{Sec-4}'s explanation of the nontrivial way in which this timestandard is in practice to be interpreted makes it LMB-CA.    

\mbox{ }  

\noindent To proceed to more complicated examples, and further PoT facets, we need the notion of configuration space $\fQ$.  
This is the set of possible values that can be taken by the configurations $Q^{\sfA}$ of one's theory, e.g. relative particle positions 
for relational particle mechanics (RPM) or spatial 3-metrics for geometrodynamics.   
Moreover, these are examples of configuration spaces with {\sl redundancies}, which play a substantial role in the following.  

\mbox{ }

\noindent 2) {\bf Configurational Relationalism} \cite{BB82, RWR, Lan, FileR, ARel}.  
Here, one subjects a configuration space $\fQ$ to some group $G$ of transformations that are held to be physically irrelevant.  
Physics is presented in this way for mathematical convenience (it is usually hard to work on the quotient space $\widetilde{\fQ} := \fQ/G$ 
that is free of such redundancy) and covers both spatial relationalism and the internal groups of irrelevant transformations of ordinary Gauge Theory.
The one builds one's action out of not $\d{Q}^{\sfA}$ but $\d_gQ^{\sfA} := \d{Q}^{\sfA} - \stackrel{\rightarrow}{G}_{\d{g}}Q^{\sfA}$ 
(for this Paper's examples, the group action $\stackrel{\rightarrow}{G}$ on the $Q^{\sfA}$ themselves cancels out in the action). 
For instance, for RPM's, $G$ consists of translations, rotations and dilations (or some subset of these due to whether scale itself is absolute or 
relative and observing that removing the translations is mathematically trivial \cite{FileR}).
For geometrodynamics, $G$ consists of 3-diffeomorphisms.\footnote{As regards other possibilities, 
one can additionally include plain or global-volume-preserving conformal transformations \cite{57}. 
$G$ = id fails for GR \cite{58}, but is consistent in the case of the strong-coupled limit of gravity \cite{San}.} 
Whilst the auxiliary variable here is usually taken to be the shift multiplier coordinate \cite{BSW}, it breaks manifest parametrization irrelevance, 
so it is replaced by the cyclic velocity $\dot{\mF}^{\mu}$ of a new `grid' variable \cite{FEPI}.  
Then an archetypal RPM action is \footnote{$\rho^{i\mu}$ are mass-weighted Jacobi inter-particle cluster relative coordinates with conjugate momenta $\pi_{i\mu}$; 
these are the most convenient relative coordinates due to their diagonalizing the RPM's kinetic term.  
The lower-case Latin letters are relative particle (cluster) labels running from 1 to $N$ -- 1 for $N$ the number of particles, 
and the lower-case Greek letter are spatial indices.}  
\beq
S = \sqrt{2}\int\sqrt{E - V}\d s \mbox{ } , \mbox{ } \mbox{ } \d s := ||\dot{\brho} - \dot{\bB} \cr \brho||^2  \mbox{ } ,
\label{Uuno}
\eeq
and the action for GR in relational form is \cite{RWR, San}\footnote{For geometrodynamics, 
the spatial topology $\Sigma$ is fixed and taken for simplicity to be compact without boundary.   
$h_{\mu\nu}$ is then a spatial 3-metric thereupon, with determinant $h$, covariant derivative $D_{\mu}$, Ricci scalar R and conjugate momentum $\pi^{\mu\nu}$.  
$\pounds_{\dot{\sF}}$ is the Lie derivative with respect to $\dot{\mF}^{\mu}$.  
%
%
$\Lambda$ is the cosmological constant.
The GR configuration space metric is ${\fM}^{\mu\nu\rho\sigma} := \{h^{\mu\rho}h^{\nu\sigma} - h^{\mu\nu}h^{\rho\sigma}\}$, i.e. 
the undensitized inverse DeWitt supermetric \cite{DeWitt67} with determinant ${\fM}$ and inverse ${\fN}_{\mu\nu\rho\sigma}$ that is itself 
the undensitized DeWitt supermetric, $h_{\mu\rho}h_{\nu\sigma} - h_{\mu\nu}h_{\rho\sigma}/2$. 
The densitized versions are $\sqrt{h}$ times the former and $1/\sqrt{h}$ times the latter. 
To represent this as a configuration space metric (i.e. with just two indices, and downstairs), one uses DeWitt's 2-index to 1-index map \cite{DeWitt67}.} 
\beq
S = \sqrt{2}\int\d^3x\sqrt{h}\int\sqrt{\mbox{R} - 2\Lambda}\,\d s \mbox{ } , \mbox{ } \mbox{ } \d s := ||\dot{\bh} - \pounds_{\dot{\sF}}\bh||_{\sbfM}^2 /2 \mbox{ } .
\eeq
These are both of the general form
\beq
S = \sqrt{2} \int_{\sN\sO\sS} \d \Omega_{\sN\sO\sS} \int \sqrt{W}||\d_{g}\bQ||_{{\mbox{\scriptsize \boldmath{${\cal M}$}}}} \mbox{ } ,  
\eeq
for NOS the notion of space in question ($\Sigma$ for geometrodynamics, trivial for RPM's), $W = E - V$ (mechanics) or $W = R - 2\Lambda$ (GR) and 
$||\d_{g}\bQ||_{\mbox{\scriptsize \boldmath{${\cal M}$}}}$ the $G$-corrected kinetic line element with configuration space metric $\mbox{\boldmath{${\cal M}$}}$.

Then one gets as a primary constraint an energy equation for the RPM and the Hamiltonian constraint 

\noindent ${\cal H} := \fN_{\mu\nu\rho\sigma}\pi^{\mu\nu}\pi^{\rho\sigma}/\sqrt{h} - \{R - 2\Lambda\}\sqrt{h} = 0$ for GR.
Also, variation with respect to the $G$-auxiliaries produces linear constraints, Lin$_{\sfZ}$, e.g. ${\cal L} := 
\sum_i \brho^i \cr \bpi_i = 0$ (zero total angular momentum for the whole universe, from varying with respect to the rotational auxiliary $\mB^{\mu}$) 
for the given RPM, or the GR momentum constraint ${\cal M}_{\mu} := - 2D_{\nu}{\pi^{\nu}}_{\mu} = 0$ from varying with respect to ${\mF}^{\mu}$.   
Solving the GR momentum constraint at the {\sl Lagrangian} level for the $\dot{\mF}^{\mu}$ is the {\it Thin Sandwich Problem} \cite{BSW, W63}. 
This is traditionally taken to be a second PoT facet \cite{Kuchar92I93}, though I have generalized this facet a) to 
the arbitrary-$G$ case of this extremization/reduction procedure: {\it Barbour's Best Matching Problem} 
\cite{BB82, RWR, Lan, B11GrybThPooley, FileR, ARel}, and b) to Configurational Relationalism itself:  
applied at {\sl whatever} (rather than the Lagrangian) \cite{APoT2}.   
\noindent Since the RPM's in 1- and 2-$d$ \cite{FileR} do have this extremization resolved,\footnote{This follows from 1) Kendall's work \cite{53} 
in the statistical theory of shapes as applied to mechanics in \cite{54, FileR} and 2) forming cones over this \cite{55, FileR}, 
which is a construct that also occurs in Molecular Physics \cite{56}.}
they serve as examples of how to resolve the PoT in approaches that eliminate Configurational Relationalism right at the outset.

Then in the presence of Configurational Relationalism, the emergent JBB time becomes subjected to the above extremization becoming attainable, 
\beq
t^{\sJ\sB\sB} = t^{\sJ\sB\sB}(0) + \stackrel{\mbox{\scriptsize extremum}}
                                                  {\mbox{\scriptsize g \, $\in$ \, $G$ \mbox{ }  of $S$}  }                                                              
\left(                                                              
\int||\d_{g}\bQ||_{{\mbox{\scriptsize \boldmath{${\cal M}$}}}}/\sqrt{2W} 
\right)  \mbox{ } .  
\eeq
It remains a simplifier for the momenta and dynamical equations of the system; 
in the GR case, $t^{\sJ\sB\sB}$ amounts to the relational recovery of proper and cosmic times in suitable contexts.  
Overall, solving the Best Matching Problem facet of the PoT once and for all at the classical level gives one an explicit  
expression for the emergent JBB time that solves the classical Frozen Formalism Problem.

\mbox{ } 

\noindent 3) {\bf The Classical Problem of Observables/Beables}.  
{\it Classical Dirac Observables} \cite{OldDir} are functionals 

\noindent D = F[$Q^{\sfA}$, $P_{\sfA}$] that Poisson-bracket-commute with 
all of a theory's constraints, whereas {\it classical \K observables} \cite{Kuchar9399BF08}  
\noindent K = F[$Q^{\sfA}$, $P_{\sfA}$] do so just with the theory's linear constraints, Lin$_{\sfZ}$.  
Bell  \cite{Bell} introduced the name and concept of `beables' as more appropriate than `observables' in the context of whole-universe Cosmology.

The Problem of Observables/Beables is then that it is hard to construct a set of these, particularly for gravitational theory.  
Clearly \K beables are more straightforward to construct (though opinions differ as to whether these are sufficient to resolve this Problem). 
Note also that Rovelli's {\it partial observables} \cite{Rovellibook, Rfqxi} do not require {\sl any} commutation, thus giving an 
approach with even less requirements.

Now, a second consequence of solving the Best Matching Problem is that this very straightforwardly implies having a full set 
of \K beables: functionals K = F[$\widetilde{Q}^{\sfA}$, $\widetilde{P}_{\sfA}$] (for $\widetilde{Q}_{\sfA}$ coordinates on $\widetilde{\fQ}$).  
On the other hand, Halliwell \cite{H03H09} has shown how to construct objects that Poisson-brackets-commute with the quadratic constraint in 
theories with trivial $G$, and I have extended this construction to nontrivial-$G$ 1- and 2-$d$ RPM examples \cite{AHall}. 
Thus \K or Dirac resolutions of the classical Problem of Beables facet of the PoT ensue.

Three further PoT Facets at the classical level -- 
the Constraint Closure Problem \cite{APoT2},
the Foliation Dependence Problem \cite{Kuchar92I93, APoT2} 
and the Spacetime Reconstruction Problem \cite{W68, APoT2} -- can all be overcome by the nature of the algebraic structure of the GR constraints \cite{58, 59, RWR, Phan, 60, APoT2}.  
The last of the three is overcome via part of the Barbour-type relational program itself. 
Moreover the three are also absent, trivial and trivial for the 1- and 2-$d$ RPM models.  
\noindent The obviously-named Global PoT facet \cite{Kuchar92I93} is classically present but can be taken to concern how time 
is but a coordinate in GR and coordinates are but in general locally valid  - on charts - and we moreover know how to mesh between charts.  
Patching GLET's together is one way of addressing part of this.
\noindent The final PoT facet - the Multiple Choice Problem \cite{Kuchar92I93} - is purely quantum-mechanical.  
Thus, overall, the classical PoT is resolved for 1- and 2-$d$ RPM's and, modulo the Thin Sandwich Problem \cite{W63}, 
for GR-as-geometrodynamics, by a combination of the Barbour-type relational approach and the benevolence of the constraint algebras in question.

\mbox{ }

\noindent At the quantum level, the PoT has further roots in the conflict of definitions between `time' in each of GR and ordinary 
Quantum Theory \cite{Kuchar92I93, Kieferbook, APoT, Rovellibook, FileR, APoT2}.  
This incompatibility complicates replace these two branches with a single 
framework in situations in which both apply, such as in black holes or the very early universe.  
Now, the purely quadratic constraint $H$ becomes 
\beq
\widehat{H}\Psi = 0
\label{FFP}
\eeq
(as opposed to e.g. a time-dependent Schr\"{o}dinger equation
\beq
\widehat{H}\Psi = i\pa_t\Psi
\eeq
for some notion of time $t$); the GR case of (\ref{FFP}) is the Wheeler-DeWitt equation \cite{DeWitt67, W68}.
This is the quantum-level Frozen Formalism Problem, and an apparent disaster occurs: 
using $t^{\sJ\sB\sB}$ fails to unfreeze this equation, so the classical resolution of this facet cannot continue.

A striking resolution, however, is that the emergent semiclassical time approach produces a new time that is both 
approximately aligned with $t^{\sJ\sB\sB}$ and itself as much of an LMB-CA implementation of Machian time as $t^{\sJ\sB\sB}$ is \cite{ACos2, FileR}.  
I.e, a somewhat more `bottom up' recovery of the classical notion; 
semiclassicality suffices for e.g. many quantum-cosmological applications such as the possible quantum origin of microwave background inhomogeneities and galaxies \cite{HallHaw}.
This semiclassical approach requires both a WKB ansatz for the wavefunction and for 3B) to hold (hence the compromise we made).

The Configurational Relationalism facet of the PoT remains solved; to some extent one likewise has quantum \K beables, 
and Halliwell has provided a {\sl separate} method for constructing semiclassical quantities that commute with $\widehat{H}$ \cite{H03H09}; 
I have again combined the two methods for 1- and 2-$d$ RPM's \cite{AHall, QuadII, FileR}.

The four further PoT facets at the quantum level are as follows. 
At the QM level, we do not know how to replace the Dirac algebra's trident of resolutions: 
one expects a Functional Evolution Problem \cite{Kuchar92I93} (QM-level version of the Constraints Closure Problem), 
a QM Foliation Dependence Problem and difficulties with Spacetime Reconstruction at the QM level. 
(The last of these is necessary due to the dynamical object being 3-geometries, and the QM fluctuations of this no longer all 
being embeddable into a single notion of spacetime \cite{W68}.)  
1- and 2-$d$ RPM's themselves fortunately remain free of these in parallel to the classical resolution \cite{FileR}.  
The Global PoT is altogether harder at the QM level: how does one mesh together {\sl unitary evolutions}? 
The Multiple Choice Problem also now appears, and is responsible for my `to some extent' clause; further work is needed on these two facets.

\section{Using $t^{\sJ\sB\sB}$ in practice: which changes are relevant?}\label{Sec-4}

\noindent To fulfil the true content of LMB-CA, all change is given opportunity to contribute to the time. 
However only changes that do so to within the desired accuracy are actually kept. 
Making approximations at the level of the terms in the formula (\ref{Uuno}) fails to capture the relevant physics approximations. 
E.g consider the Earth-Sun-Andromeda system.
Whilst $V$ due to Andromeda is {\sl felt} by solar system objects, what counts is that is felt {\sl extremely evenly} by them  
all, so it has no appreciable role in the physics of the solar system itself.  
[For $V$, Andromeda being far is offset by it being massive ---   
$m_{\sA}/\rho_{\sE\sA}$ versus $m_{\sS}/\rho_{\sE\sS}$.
But in the equations of motion, one is to compare tidal-type terms $m_{\sA}/\rho_{\sE\sA}^3$ versus $m_{\sS}/\rho_{\sE\sS}^3$,  
so Andromeda being far now greatly dominates it being massive.] 
Thus one must get the equations of motion from (\ref{Uuno}) and then assess approximations there.

\mbox{ }

\noindent I argue against Barbour's terming $t^{\sJ\sB\sB}$ ``the'' ephemeris time because ephemeris time is not a unique entity, 
nor does the method he presents entail any of the specific properties of any iterative procedure used to compute an ephemeris time \cite{AlmaSup}.  
%
%
The relational underpinning itself does nothing new for the actual physical calculation of the ephemeris time, whereas the {\sl practical} knowledge of how to 
handle the solar system physics iteratively to set up the ephemeris, which was developed without ever any reference to relationalism, is fully used. 
In particular, the formula for $t^{\sJ\sB\sB}$ is {\sl not} directly implementable (even in cases for which its extremization is 
explicitly solvable for, itself another problem with any practical use of that formula).  
On the other hand, the relationalism does provide an alternative philosophical underpinning by which the 
astronomical timestandard calculation can indeed be said to rest on a relational rather than absolute view of the universe.  

\mbox{ } 

%
%
\noindent Finally, `heavy' degrees of freedom contribute to the zeroth-order estimate for $t^{\sJ\sB\sB}$ \cite{SemiclI} in the quantum-cosmological setting. 
While this might look more like AMR than LMB(-CA), I subsequently show in \cite{ACos2} that 
it can be taken as the zeroth order part of a perturbation scheme which does give whatever other degrees of freedom opportunity to contribute 
to the time at first order.

\section{`Marching in step' criteria?}\label{Sec-5}

In the Newtonian picture, all good clocks {\sl happen} to march in step with absolute time.  
A stronger position on this concerns how clock manufacturers' antecedents separately agreed with {\sl natural standards} 
rather than merely with each other.  
I.e. they had to produce {\sl useful devices} where the measure of that usefulness was marching reasonably well in step 
with solar system phenomena such as night and day, where the Sun appears in the sky, or the length of the seasons.  
Clock designs failing to march in step with such things would be deemed to be poor and selected against by people 
wishing to keep successively more accurate appointments. 
Thus the agreements to reasonable approximation between all surviving types of clocks have a clear mechanism behind them. 
In the Newtonian picture, one might tie this up with some kind of postulate that the celestial 
bodies move in very close accord with absolute time.

Barbour then suggested \cite{Bfqxi} that in the relational LMB picture, rather, good clocks are `marching in step' with the emergent JBB time.
He argued furthermore that this criterion applies universally, and uniquely to this timefunction.

\mbox{ }

Unfortunately, Barbour's term `marching is step' also carries connotations of a concept that is far more often called  
{\it synchronicity} \cite{Jammerteneity}.
This matters because the mechanics argument Barbour gives carries straight over to the GR case, for which it {\sl cannot} hold by how 
relativistic synchronization requires a procedure rather than just occurring naturally.   
Alerted thus, it is straightforward to spot that Barbour's argument contains a theory-independent circularity \cite{FileR} 
(he substitutes the quadratic constraint into a rearrangement of itself).

Thus the universal basis for marching in step is lost, so one is left having to consider a procedure for approximately patching 
disparate observers' GLET's together.

\section{Patching GLET's together}\label{Sec-6}

\noindent Consider modelling two quasi-isolated island subsystems within a universe.
Then Sec \ref{Sec-4}'s approximation applied to each will ensure that the details of the other's contents will contribute negligibly. 
Thus each's timestandard constructed as a GLET would be independent of the other's.
Thus ephemeris time type constructions do not in practice by themselves appear to provide a common timestandard.   
[There {\sl is} a JBB timefunction that interpolates between the two; however, neither observer would know enough about the other's local island 
subsystem so as to be able to at all accurately compute that interpolatory function.]  

\mbox{ }  

A more relevant line of enquiry may be how observers in different places will have access to  
different changes. 
Whilst the GLET's these establish will not in general march in step with each other, 
the GLET concept has a physically natural means of approximate patching via examination 
of those changes in the universe that can be observed by both of the observers.
For, by each choosing a time that works well for the STLRC observed by each observer, both times work well 
to describe the mutually-observed change, so the two timestandards are reasonably accurately synchronized.
(This is up to a linear transformation that gives each observer freedom in choice of time-unit and of `calendar year zero'.) 
There are two strengths of sharing: of the same subsystem versus of the same {\sl type} of subsystem 
(for types that have sufficiently universal physics, in parallel to `standard candles').  
Pulsars might often serve as shared observable subsystems, due to their long-range detectability as well as longevity and regularity.  
Cosmological observations might also serve, 
though only very inaccurate timestandards ensue from them (at least for observers with our technology and history).   
Two comments are in order.

\mbox{ } 

\noindent  1) Unlike the {\sl incontestability} of LMB by including all changes, practical LMB-CA timestandards involves 
may still be contestable by external physics; e.g. if the Solar System does not keep time with the pulsars, 
which do we consider to be at fault?  

\noindent 2) Patching together GLET's may well be of relevance to travel and settlements in space.

\section{Conclusion: ties to Problem of Time approaches} \label{Sec-7}

To look into resolving the Problem of Time (PoT), it greatly helps to have the deepest possible understanding of time itself.   
This Paper suggests time is conceptually a generalized local ephemeris time (GLET) based on a sufficient totality of locally relevant change (STLRC) --- 
a particular interpretation of Mach's `time is to be abstracted from change', and one that holds even truer to the Astronomy that inspired Barbour's own whole-universe interpretation.  
I portrayed emergent Jacobi-Barbour-Bertotti time $t^{\sJ\sB\sB}$ as a GLET and as a resolver of the classical PoT modulo one being able to 
solve the Best Matching Problem for whichever specific case one is investigating; a further bonus of solving this is that \K beables
are then known.
Let us finally use the GLET concept as a selector and remoulder of a more general set of PoT strategies: does one seek to find 
or append a time prior to quantization, to obtain an emergent time after quantization, to take the timelessness at face value 
or to consider the notion of history to supplant that of time (Histories Theory)?  
Below is an approach favoured by this paper's considerations {\sl for}.\footnote{Time being a GLET based on a STLRC can be used as a sweeping argument {\sl against}  each of 
the following approaches presented in the reviews \cite{Kuchar92I93}.
a) hidden times (hidden within the general classical theory),
b) the matter times (obtained by appending matter that provides a time), and 
c) unimodular time (from a dynamically-arising cosmological constant) of `time before quantization' schemes \cite{Kuchar92I93, APoT}. 
This is because these are all abstracted from {\sl particular} changes rather than giving all changes an opportunity to contribute.  
Case-by-case arguments against each of these are given in \cite{APoT2, FileR}; on some occasions [e.g. some {\it reference fluids} examples of b)] the matter in question is 
not only appended for the purpose of having a time but also intangible, giving it rather too many parallels to absolute time for comfort.  

Finally, a lot of uses of the word `relational' in Ashtekar variables/Loop Quantum Gravity \cite{Rovellibook} are, 
rather of the AMR `any' form, suggesting far from maximal exploitation of the full meaning of `relational' hitherto in that literature.  
Thus the present work should unleash quite a conceptual revolution in that subject too.}  

\mbox{ } 

\noindent The `time after quantization' semiclassical approach very nicely embodies GLET; 
it is quite, but not perfectly, aligned with \cite{SemiclI, ACos2, FileR} the classical $t^{\sJ\sB\sB}$.  
There is \cite{ACos2} a semiclassical parallel of the classical perturbative scheme mentioned in Sec \ref{Sec-4} that is likewise LMB-CA; 
the timestandard it produces is, to high accuracy not quite the same as $t^{\sJ\sB\sB}$, which is to be expected since 
{\sl quantum} changes now contribute.  

\mbox{ }

\noindent Many approaches that take the timelessness of (\ref{FFP}) at face value at the primary level have subsequent semblances of time that are based 
on the `any' implementation \cite{PW83GPP,Page1} (or Barbour's `all' implementation \cite{B94II, EOT}). 
I already argued how much of Barbour's records scheme could be recast in terms of local records \cite{Records, FileR}; 
now also the associated semblance of dynamics scheme would likewise be recast as local.  
Thus this sort of scheme survives, as does the \CPI \cite{PW83GPP} where the conditioning now 
ends up being  preferentially based on the GLET rather than just `any' clock variable.  
\cite{FileR} contains a tabulation of timeless strategies that allow for each of `any', `all' and 
`STLRC' interpretations for each's semblance of time (thus roughly multiplying the existing count of such strategies by a factor of 3).  
Histories Theory \cite{Hartle} can also be rendered compatible with GLET.  


\noindent The semiclassical approach is foundationally unsatisfactory on other grounds: its crucial WKB assumption 
is not internally justified in this scheme \cite{Kuchar92I93}.
\noindent However, combined histories-records-semiclassical approaches \cite{H99, H03H09, AHall, FileR} do better in this regard.  
Schemes of this nature can indeed also be taken to live within the STLRC interpretation of Mach's `time is abstracted from change'.  
One such combined scheme is the setting for Halliwell's \cite{H03H09}; this is also precisely the same setting as that I mentioned 
at the end of Sec \ref{Sec-3} as regards promoting 1- and 2-$d$ RPM quantum \K beables to semiclassical quantum Dirac beables.
Thus, combined with Endnotes 5-6), this accounts for 6/8ths of the PoT for the RPM models in question.  

\mbox{ } 

\noindent {\bf Acknowledgements}: I thank Jeremy Butterfield and Alexis de Saint-Ours for comments. 
I was funded by a grant from the Foundational Questions Institute (FQXi) Fund, a donor-advised fund of the Silicon Valley Community Foundation 
on the basis of proposal FQXi-RFP3-1101 to the FQXi.  
I thank also Theiss Research and the CNRS for administering this grant.


\end{document}